\documentclass[12pt]{article}
\usepackage{amsmath}
\begin{document}
\centerline{\bf Twists of quantum groups and noncommutative field theory
\footnote{This research was supported in part by the grants 
RFBR 05-01-00922 and CRDF RUM-1-2622-ST-04}} 

\medskip

\centerline{P.~P.~Kulish}

\centerline{{\it St.Petersburg Department of Steklov Institute of Mathematics,}} 

\centerline{{\it Helsinki Institute of Physics}}
 
\centerline{{\it and}}

\centerline{{\it High Energy Physics Division, Department of Physical Sciences, University of Helsinki}}

\medskip

\begin{abstract}
The role of quantum universal enveloping algebras of symmetries
in constructing a noncommutative geometry of space-time
and corresponding field theory is discussed. It is shown that
in the framework of the twist theory of quantum groups,
the noncommutative (super)space-time defined by coordinates
with Heisenberg commutation relations, is (super)Poincar\'e invariant,
as well as the corresponding field theory. Noncommutative parameters 
of global transformations are introduced.
\end{abstract}

One of attempts to study the structure of space-time at Planck scale is
related with a possible noncommutative nature of space-time, hence, with a
noncommutative geometry (for references see e.g. \cite{AW,A-G}). 
In this paper we would like to draw attention to interrelations between
noncommutative quantum field theories and quantum groups~\cite{CK}. 
Recently, an active research takes place in
noncommutative field theory related to noncommutative geometry (see
the reviews \cite{DN,Sz} and references therein). One source of
examples of noncommutative geometry is the theory of quantum
groups~\cite{Dr1,FRT,ChP}. The reason for this is that the latter
are, loosely speaking, deformations of Lie groups, which provide numerous
geometric structures. There are corresponding structures in quantum
groups (QG), where the commutative algebra of functions $F(G)$ on a Lie
group $G$ is deformed into an appropriate noncommutative algebra
$F_q(G)$, which is  defined e.g.\ by generators and relations \cite{FRT}. 
Homogeneous spaces are also subject to deformation, for example
$SL(2) \rightarrow SL_q(2)$ or $SU(2) \rightarrow SU_q(2)$
and two-dimensional plane $(x,y)\rightarrow$
``quantum plane'' $(x,y)_q$, or two dimensional sphere 
to the Podles $q$-sphere $(x,y,z)\rightarrow (x,y,z)_q.$
It has been observed by several authors (see e.g.\ \cite{CK,KM,O,M})
that the twist theory of quantum groups provides a very useful tool
for constructing non-commutative geometry of space-time, including vector 
bundles, measure, and equations of motion and their solutions. 

The most important space of relativistic theory is four-dimensional
Minkowski space-time ${\mathcal M}$, with coordinates $x^\mu$, and
with the Poincar\'e algebra acting on $x^\mu$.
To construct NC field theory, the commutative algebra of functions
$C({\mathcal M})$ on ${\mathcal M}$ is deformed to a noncommutative
(NC) algebra $C_\theta ({\mathcal M})$. This algebra is generated
by NC coordinates $x^\mu$,
and probably the simplest relations among the $x^\mu$ are
\begin{equation}\label{cr01}
[\hat{x}^\mu, \hat{x}^\nu] = \hat{x}^\mu \hat{x}^\nu -
\hat{x}^\nu \hat{x}^\mu= i\theta^{\mu \nu},
\end{equation}
with a constant antisymmetric matrix $\theta$ (see \cite{DN,Sz}).

There are many possible commutation relations (CR) for $x^\mu, x^\nu$
with the right hand side linear or quadratic in $x^\mu$
(see \cite{W,LW}). However, those written above follow from a special
limit of string theory \cite{SW} and have attracted substantial interest.

To construct a field theory on noncommutative space-time with
CR (\ref{cr01}) for the coordinates,  one has to substitute the commutative
algebra of fields (functions on ${\mathcal M}$) by the noncommutative
algebra $C_\theta$. In the case of the CR (\ref{cr01}) there is a Weyl-Moyal
correspondence between these algebras through the Fourier transform. It
maps a smooth function $\varphi(x) \in C({\mathcal M})$ to
an element of the algebra $C_\theta$,
\begin{equation}
\label{ft}
\varphi(\hat{x})=\frac{1}{(2\pi)^4}\int d^4k \tilde{\varphi}(k)
\exp({ik\hat{x}}),
\end{equation}
with $\tilde{\varphi}(k)$ being the Fourier transform of the function
$\varphi(x)$, and $k\hat{x}  = k_{\mu}\hat{x}^\mu$
\[
\tilde{\varphi}(k) = \int d^4x \varphi(x)e^{-ikx}.
\]
Then the noncommutative product in the algebra $C_\theta$ is
\begin{align}
\label{ncp}
\varphi(\hat{x})g(\hat{x})&=
\int \frac{dk_1}{(2\pi)^4} \frac{dk_2}{(2\pi)^4} \tilde{\varphi}(k_1)
\tilde{g} (k_2) e^{ik_1\hat{x}} e^{ik_2\hat{x}}  \nonumber\\[3pt]
&= \int \frac{dk_1}{(2\pi)^4} \frac{dk'_2}{(2\pi)^4} \tilde{\varphi}(k_1)
\tilde{g} (k'_2-k_1) e^{-i \theta(k_1, k'_2)} e^{ik'_2 \hat{x}},
\end{align}
where the notation
$\theta(k,p):= \frac{1}{2} \theta^{\mu \nu} k_\mu p_\nu$
is introduced for the antisymmetric quadratic form.

Interpreting\, the convolution of\,\, $\tilde{\varphi}(k_1)$\,\, and
\,\,$\tilde{g}(k_2)$\,\,
with\, the weight\, function\hfill\break
$\exp (-i\theta(k_1, k_2))$ as the Fourier
transform of a new product ($\ast$-product) of the elements \hfill\break
$\varphi(x),\, g(x) \in C({\mathcal M})$ one gets
\begin{equation}\label{newpr}
\varphi(x) \ast g(x) =
\int \frac{dk_1}{(2\pi)^4} \frac{dk'_2}{(2\pi)^4} \tilde{\varphi}(k_1)
\tilde{g}(k'_2-k_1) \sum_n \frac{1}{n!} (-i\theta(k_1, k'_2))^n
e^{ik'_2x}.
\end{equation}
It is not difficult to check that this $\ast$-product on
$C({\mathcal M})$ is still associative, albeit noncommutative.
The exponential function $\exp({ik\hat{x}})$ generates symmetrized
$\ast$-pro\-ducts of $\hat{x}^\nu$, which coincide
with the usual products of commutative $x^\nu$. 
Let us point out that the ``$\ast$-product" is a general notion of 
deformation quantization (see the review \cite{St}).

It follows that a field theory on the NC space-time can be constructed using
fields $\varphi(x) \in C({\mathcal M})$, but with multiplication given by
the $\ast$-product. To fix
an action one needs a linear functional on $C_\theta$, and it is
represented as an integral on $C({\mathcal M})$ of the usual form, e.g.
\begin{equation}\label{ac}
S[\varphi] = \int dx \{\frac{1}{2} (\partial_\mu \varphi(x))^2 -
\frac{m^2}{2} (\varphi(x))^2 - \frac{\lambda^2}{4!} (\varphi(x))^4_\ast\}.
\end{equation}
The integral of the $\ast$-product of several functions is invariant
only under the cyclic permutations, similarly to the trace of operators:
\[
\int\, dx f_1(x) \ast f_2(x) \ast \ldots \ast f_n(x)= \int dx\, f_2(x) \ast
\ldots \ast f_n(x) \ast f_1(x).
\]
With the $\ast$-product chosen (\ref{newpr}) one has 
$\int dx f_1(x) \ast f_2(x) = \int dx f_1(x) \cdot
f_2(x)$, and NC field theory and ordinary field theory coincide on
the free field level (the  action with quadratic terms only). However, the
interaction term being written as the $\ast$-product of the fields,
describes a nonlocal interaction, e.g.\ for the
$\varphi^3_{\ast}$-theory
\[
\begin{split}
\int dx\, (\varphi(x))^3_\ast & = \int \prod_a
\left(\frac{dk_a}{(2\pi)^4} \tilde{\varphi}(k_a)\right)
\exp (-i\sum_{b<c}\theta (k_b, k_c))\delta (\sum_j k_j) \\
& = \int \prod_a dx_a\,\varphi(x_1)\varphi(x_2)\varphi(x_3)
\exp (2i(x_1-x_3) \theta^{-1} (x_2-x_3)),
\end{split}
\]
provided that the matrix $\theta$ is invertible (or one has to
restrict the arguments to those $x^j$ for which $\theta^{ij}$ has
an inverse).

Quantization of the scalar field theory with the action $S[\varphi]$
by path integral methods yields the standard perturbation theory, but the
interaction vertices include an extra oscillating factor,
\[
V(k_1, \ldots, k_4)= \frac{\lambda^2}{4!} \delta (\sum_a k_a) \prod_{b<c}
e^{-\frac{i}{2}\theta_{\mu \nu} k^\mu_b k^\nu_c}.
\]
This factor has only cyclic symmetry (due to the delta-function) and
results in different contributions as compared to local QFT, and even in
a different structure of the Feynman diagrams (planar versus non-planar
graphs). The diagrammatic analysis of unitarity yields a condition on
$\theta^{\mu \nu}$: $\theta^{0 j} = 0$. Thus the time coordinate
commutes  with the space coordinates, and
one can apply the Hamiltonian formalism for the action (\ref{ac}).

 Reformulating NC space-time field theory as a usual one (\ref{ac})
with a nonlocal interaction,
it is possible to apply standard techniques to quantize it.
An obvious drawback is the appearence of the set of constants
$\theta^{\mu \nu}$ breaking the Lorentz invariance:
$x^{\mu}\rightarrow \Lambda^{\mu}_{\,\,\nu}x^{\nu},$
$\theta^{\mu \nu}\rightarrow \Lambda^{\mu}_{\,\,\alpha}
\Lambda^{\nu}_{\,\,\beta}\,\theta^{\alpha \beta}=
\widetilde{\theta}^{\mu \nu}\neq\theta^{\mu \nu}.$ To cure this problem
we propose to use a quantum group technique.

In this discussion we need such objects from the theory of quantum groups
as a Hopf algebra ${\mathcal H}$, its ${\mathcal H}$-module
algebra ${\mathcal A}$, ${\mathcal H}$-modules and
${\mathcal A}$-modules $V, W$ (linear spaces for ${\mathcal H}$- and
${\mathcal A}$-representations). At the same time these objects have
a physical interpretation: ${\mathcal H}$ is the symmetry algebra of
the system under consideration, ${\mathcal A}$ is the algebra of
observables, and their representation space is the space of states of
the system. There are also additional structures, such as a $\ast$-operation
(real form), a scalar product etc., which will be introduced later.

 The symmetry of the relativistic field theory is described by
the universal enveloping algebra ${\mathcal U}({\mathcal P})$ of the
Poincar\'e Lie algebra ${\mathcal P}$ with generators of
translations  $P_\mu$ and rotations $M_{\mu \nu}$:
\begin{align}
[P_\mu, P_\nu] & = 0, \nonumber\\[4pt]
[M_{\mu \nu}, M_{\alpha\beta}] & = -i (\eta_{\mu\alpha}M_{\nu\beta}
-\eta_{\mu\beta}M_{\nu\alpha}-\eta_{\nu\alpha}M_{\mu\beta}
+\eta_{\nu\beta}M_{\mu\alpha}),\label{Pcr}\\[4pt]
[M_{\mu \nu}, P_\alpha] & = -i (\eta_{\mu\alpha} P_\nu -
\eta_{\nu\alpha}P_\mu).\nonumber
\end{align}
The essential part of the Hopf algebra structures
${\mathcal H}(m, \Delta, \gamma, \epsilon)$
(see \cite{ChP,Dr1} for details) is given by the associative product
(with the commutation relations (\ref{Pcr})
for ${\mathcal U}({\mathcal P})$ in our case) and by a coproduct map
$\Delta: {\mathcal H}\rightarrow {\mathcal H}\otimes {\mathcal H}$
defining an action
of the Hopf algebra ${\mathcal H}$ in the tensor product of two (or
more) of its representations. The action of the
generators $Y \in {\mathcal P}$ in a tensor product $V\otimes W$ is
given by the symmetric map (coproduct)
$\Delta (Y)=Y\otimes 1 + 1 \otimes Y$, or
\begin{equation}\label{prcop}
\Delta(Y) (v \otimes w) = (\hat{Y}v) \otimes w + v\otimes (\hat{Y}w),
\end{equation}
where the hat means the action of a Hopf algebra element in the
corresponding representation space. There are two other maps in
the definition of the Hopf algebra: the counit $\epsilon: {\mathcal H}
\to {\mathcal C}$ (a one-dimensional representation of ${\mathcal H}$) and
the antipode $\gamma : {\mathcal H} \to {\mathcal H}$,
which is an algebra antihomomorphism.
These maps are subject to quite a few axioms, of course \cite{Dr1,FRT,ChP}.
On the generators of  ${\mathcal U}({\mathcal P})$ the antipode and counit are:
$\gamma (Y) = - Y,\, \epsilon(Y) = 0, \epsilon(1) = 1$.

There is a useful transformation (twist) of the structure maps of a Hopf
algebra, which is an equivalence relation among Hopf algebras,
preserving their category of representations. This transformation
${\mathcal H}\rightarrow {\mathcal H}_t$ is realized by an invertible
twist element \cite{Dr2} 
\[
{\mathcal F}=\sum_i f^i_1 \otimes f^i_2 
\in {\mathcal H}\otimes {\mathcal H}.
\]
It does not change 
the multiplication in ${\mathcal H}$, but transforms the coproduct according to
\[
\Delta (h) \rightarrow \Delta_t(h) = {\mathcal F} \Delta (h){\mathcal
  F}^{-1}, \quad h \in {\mathcal H}.
\]
This similarity transformation preserves the coassociativity of the
twisted coproduct if ${\mathcal  F}$ satisfies
the following twist equation (two-cocycle condition) in
${\mathcal H} \otimes {\mathcal H}\otimes {\mathcal H}$ \cite{Dr2}
\begin{equation}\label{TE}
{\mathcal F}_{12} (\Delta \otimes id) {\mathcal F} = {\mathcal F}_{23}
(id \otimes \Delta) {\mathcal F},\qquad
(\epsilon\otimes {id}){\mathcal F}=1\otimes 1,
\end{equation}
where ${\mathcal F}_{23}$ means $\sum_i 1 \otimes f^i_1 \otimes f^i_2
\in {\mathcal H}^{\otimes 3}$, and $(\Delta \otimes id) {\mathcal F}:
= \sum_i \Delta (f^i_1) \otimes f^i_2 \in {\mathcal H}^{\otimes 3}$.
The twist does not change the counit homomorphism, but similarity-transforms
the antipode:
\begin{equation}\label{anti}
\gamma(Y) \to \gamma_t(Y) = u\gamma(Y)u^{-1},\quad \textrm{where}\ \ \
u= \sum_i f^i_1\cdot \gamma (f^i_2) \in {\mathcal H}.
\end{equation}
Usually the twist element is not symmetric under the permutation of its tensor
factors: ${\mathcal F} \neq {\mathcal F}_{21} = \sum_i f^i_2 \otimes
f^i_1$. Hence, the twisted coproduct $\Delta_t (h): = \sum h_{(1)}
\otimes h_{(2)}$ is also non-symmetric
\[
\Delta_t (h) \neq \Delta^{op}_t (h)=\sum h_{(2)} \otimes h_{(1)}.
\]
However, for the quantum group case the coproduct $\Delta_t (h)$ 
and the opposite coproduct $\Delta^{op}_t (h)$ are related by a
similarity transformation with the ${\mathcal R}$-matrix:
\[
{\mathcal R} \Delta_t = \Delta^{op}_t {\mathcal R}, \quad {\mathcal R} =
\sum {\mathcal R}_1 \otimes {\mathcal R}_2 \in
{\mathcal H} \otimes {\mathcal H}.
\]
In our case, starting with the symmetric coproduct (\ref{prcop}) the
${\mathcal R}$-matrix is given by
${\mathcal R}={\mathcal F}_{21} {\mathcal F}^{-1}$.

There are well-known statements from the theory of quantum groups
which will be used in our discussion of a particular case of
noncommutative space-time. Having an action of ${\mathcal H}$ 
on an associative algebra
${\mathcal A}$ with consistency of the coproduct of ${\mathcal H}$ and
multiplication of ${\mathcal A}$ (a Leibniz rule),
\[
\hat{h} (a\cdot b) = \sum (\hat{h}_{(1)}a)\cdot (\hat{h}_{(2)} b),
\]
the multiplication in ${\mathcal A}$ has to be changed after twisting
${\mathcal H}\rightarrow {\mathcal H}_t$ to preserve this consistency. 
The new product in ${\mathcal A}_t$ is 
\begin{equation}\label{tp}
a \ast b = \sum (\hat{\bar{f}}^i_1 a) \cdot
(\hat{\bar{f}}^i_2 b), \quad a,b \in {\mathcal A}_t,
\end{equation}
where a notation was introduced for ${\mathcal F}^{-1}:= \sum
\bar{f}^i_1 \otimes \bar{f}^i_2$, and the action (representation) of
elements from ${\mathcal H}$ on elements from ${\mathcal A}_t$
is the same as before twisting. 

The product $\varphi(x^{\mu}_{1})\ast\varphi(x^{\nu}_{2})$ of
quantum fields with independent arguments belongs to the tensor
product of two copies of the algebra $C_\theta ({\mathcal M})$.
After twisting of ${\mathcal H}$ the elements of different copies
of ${\mathcal A}\otimes{\mathcal A}$ will not commute:
$$
(a_{1}\otimes 1)(1\otimes a_{2})=(a_{1}\otimes a_{2}), \qquad \textrm{but}
$$
\begin{equation}\label{ttp}
(1\otimes a_{2})(a_{1}\otimes 1)=(\hat{\mathcal R}_2a_{1})
\otimes(\hat{\mathcal R}_1 a_{2})\neq (a_{1}\otimes a_{2}),
\quad \forall a_{1}, a_{2}\in {\mathcal A}.
\end{equation}
(Recall that the hat indicates the action of Hopf algebra elements on the
 relevant representation spaces.) If in addition one has 
an ${\mathcal H}$-covariant representation of the algebra 
${\mathcal A}$ in a vector space $V$, then the action of the elements of 
${\mathcal A}$ on vectors of $V$ will be changed correspondingly 
$a \cdot v \to a \ast v = \sum (\hat{\bar{f}}^i_1 a) \cdot
(\hat{\bar{f}}^i_2 v)$.

It is important that real forms survive a twist.
Recall that a $*$-operation (real form) on a Hopf algebra
${\mathcal H}$ means an antilinear involutive algebra anti-au\-to\-morphism
and coalgebra automorphism. Due to the uniqueness of the antipode,
the identity $\gamma *=* \gamma^{-1}$ is always valid, and
one can re-define the real form as $\gamma^{2n} *$ for any integer number $n$.
We can also consider homomorphic and anti-cohomomorphic antilinear
operations of the kind $\xi =\gamma^{2n+1} *$.

To ensure consistency between real forms and the action of ${\mathcal H}$
on some ${\mathcal H}$-mod\-ule algebra ${\mathcal A}$ with anti-involution
$a\to \bar a$, one has to require $\overline{(h a)} = \gamma (h^*) \bar a$,
for $h\in {\mathcal H}$ and $a \in {\mathcal A}$.  So by the real form of a
quantum algebra we will mean a homomorphic and anti-cohomomorphic antilinear
involution $\xi = \gamma \circ *$.

Twisting a Hopf algebra ${\mathcal H} \to {\mathcal H}_t$ the same
$*$-operation is defined on ${\mathcal H}_t$ if the twist $\mathcal F$
satisfies the condition
\begin{equation}\label{tin1}
{\mathcal F}^*= \sum f_1^* \otimes f_2^* = {\mathcal F}^{-1} =
\sum {\bar f}_1 \otimes {\bar f}_2.
\end{equation}
For the involution $\xi$ the analogous natural requirement is \cite{KM}
\begin{equation}\label{tin2}
(\xi \otimes \xi){\mathcal F}= \tau ({\mathcal F}):= {\mathcal F}_{21} =
\sum f_2 \otimes f_1,
\end{equation}
where $\tau$ is the permutation of the factors in
${\mathcal H} \otimes {\mathcal H}$.

Suppose now that ${\mathcal A}$ possesses a measure $\mu$, i.e.\ a linear
functional positive on elements of the form  $a\cdot \bar a$ (like the
function algebra on a locally compact topological space does).
The same measure is valid for ${\mathcal A}_t$, for these
${\mathcal H}$-module algebras ${\mathcal A}$ and ${\mathcal A}_t$
coincide as linear spaces \cite{KM}.
Indeed, we find $a* \bar a = \bar f_1 a\cdot \bar f_2 \bar a =
\bar f_1 a\cdot \overline{(\xi({\bar f_2}) a)} $. If
identity (\ref{tin2}) is fulfilled, the relation
$\bar f_1\otimes \xi(\bar f_2) = \xi(\bar f_2) \otimes \bar f_1$
holds as well
and, consequently,  $\bar f_1\otimes \xi(\bar f_2)$ can be represented
by a sum $\sum \varphi_i\otimes \varphi_i$. Further, we have
$a * \bar a = \sum\varphi_i a\cdot \overline{\varphi_i a}$, and
therefore $\mu(a* \bar a)\geq 0$.
In case that (\ref{tin1}) is true, one can extend the Hopf algebra
by adding the square root of 
 the element $u$ that was introduced in (\ref{anti}).  
It is straightforward that the
composition of the coboundary twist with the element
$\Delta(u^{-\frac{1}{2}})(u^\frac{1}{2}\otimes u^\frac{1}{2})$
and successive twist with the element
$(u^{-\frac{1}{2}}\otimes u^{-\frac{1}{2}}) {\mathcal F}^{-1}
(u^\frac{1}{2}\otimes u^\frac{1}{2})$
obeys (\ref{tin2}). This double transformation is carried out
by means of the 2-cocycle $\Delta(u^{-\frac{1}{2}}){\mathcal F}^{-1}
(u^\frac{1}{2}\otimes u^\frac{1}{2})$,
and the required property (\ref{tin2}) readily follows from
(\ref{tin1}) and the identity
$(u\otimes u)\tau (\gamma \otimes \gamma)({\mathcal F}^{-1})=
{\mathcal F}\Delta(u)$
fulfilled for any solution to the twist equation \cite{Dr2}
(the element $u$ is exactly the same as the one
taking part in the definition of the twisted antipode (\ref{anti})).
So we can apply all the previous considerations
to this composite twist, which differs from initial one by
an inner automorphism only.

Let's deform the Poincar\'e algebra
${\mathcal U}({\mathcal P})$ as a Hopf algebra by a
simple twist element depending only on the generators of translations
$P_{\mu}$ (an abelian subalgebra of ${\mathcal P}$) \cite{CK}:
\begin{equation}\label{abt}
{\mathcal F}=
\exp\left(\frac{i}{2}\theta^{\mu \nu}P_{\mu}\otimes P_{\nu}\right)
\end{equation}
with a constant matrix $\theta^{\mu \nu}$ (we take it to be real and
antisymmetric).
As an associative algebra ${\mathcal U}_t({\mathcal P})$ is not changed
(we have the same commutation relations of generators $M_{\mu \nu},$
$P_{\alpha}$) nor is the coproduct of $P_{\alpha}:$
$\Delta_t(P_{\alpha})=\Delta(P_{\alpha}).$ However, the coproduct of
$M_{\mu \nu}$ is changed:
\begin{equation}\label{cop1}
\begin{split}
\Delta_t(M_{\mu \nu})=\text{Ad}
\left(\exp\left(\frac{i}{2}\theta^{\alpha \beta} P_{\alpha} \otimes P_{\beta}
\right)\right) \Delta(M_{\mu \nu})\qquad\qquad\qquad\qquad
\\ \qquad =
\Delta(M_{\mu \nu})-\frac{1}{2}\theta^{\alpha \beta}
\left((\eta_{\alpha \mu}P_{\nu}-\eta_{\alpha \nu}P_{\mu})
\otimes P_{\beta}+P_{\alpha}\otimes
(\eta_{\beta \mu}P_{\nu}-\eta_{\beta \nu}P_{\mu})\right).
\end{split}
\end{equation}
It was  already mentioned that the coproduct defines
an action of the Hopf algebra on the product of elements from
${\mathcal A},$ and the product of ${\mathcal A}$ is also changed
accordingly, to be consistent with $\Delta_t.$ The algebra
${\mathcal C}({\mathcal M})$ is generated by the $x^{\mu},$ and after
twisting ${\mathcal C}({\mathcal M})\rightarrow{\mathcal C}_t({\mathcal M})$
the new product is
\begin{equation}\label{sp1}
\begin{split}
x^{\mu}\ast x^{\nu} = \sum(\hat{\bar f}_1x^{\mu})(\hat{\bar f}_2x^{\nu})
\qquad\qquad\qquad\qquad\qquad\qquad\qquad\qquad\quad \\
=\sum_{k=0}^{\infty}\frac{(i/2)^k}{k!}\prod_{j=1}^{k}\theta^{\mu_j, \nu_j}
\left(\partial_{\mu_1}\ldots\partial_{\mu_k}x^{\mu}\right)
\left(\partial_{\nu_1}\ldots\partial_{\nu_k}x^{\nu}\right)\qquad\\[-4pt]
=x^{\mu}x^{\nu}+\frac{i}{2}\theta^{\mu \nu}.
\end{split}
\end{equation}
Hence,
\begin{equation}\label{cr02}
[x^{\mu}, x^{\nu}]_{\ast}:=x^{\mu}\ast x^{\nu} - x^{\nu}\ast x^{\mu}=
i\theta^{\mu \nu},
\end{equation}
and this yields ${\mathcal C}_t({\mathcal M})={\mathcal C}_{\theta}.$
One can check that with the deformed coproduct (\ref{cop1}) these CR
are invariant under the action of $M_{\mu \nu}$ \cite{CK}. 

One has to mention that the interpretation of the Weyl - Moyal product
using abelian twist has been known for some time in the deformation 
quantization (see~\cite{GZ,BGGS} and references  therein). 

It is possible to write explicitly the star product 
of $x^\mu$ and $\phi (x)$ according to (\ref{sp1}) 
\begin{equation}\label{sp2}
x^{\mu}\ast \phi (x) = x^{\mu} \phi (x) + 
\frac{i}{2}\theta^{\mu \nu} \partial_{\nu} \phi (x). 
\end{equation} 
Hence, the $\ast$-multiplication by $x^\mu$ can be represented as 
an action by the operator 
$x^{\mu} + \frac{i}{2}\theta^{\mu \nu} \partial_{\nu}$ in the space 
$C({\mathcal M})$. 
Similarly, the star product of $\varphi(x)$ and $g(x)$ can be 
represented as an action of a differential operator on the second 
factor 
\begin{equation}\label{sp3}
\varphi(x) \ast g(x) = \left(\varphi (x) + 
\sum_{k=1}^{\infty}\frac{(i/2)^k}{k!}
\prod_{j=1}^{k}\theta^{\mu_j, \nu_j}
\left(\partial_{\nu_1}\ldots\partial_{\nu_k}\varphi(x) \right) 
\left(\partial_{\mu_1}\ldots\partial_{\mu_k}\right) \right) g(x). 
\end{equation}
However, one can use instead of usual functions on Minkowski space-time 
with the star product the algebra generated by noncommutative $\hat{x}^\mu$. 
Correspondence between the bases of these algebras in our case (\ref{cr01}), (\ref{cr02}) 
is given by the generating functions of their bases $\exp (ikx)$ and $\exp (ik\hat{x})$ 
(cf (\ref{ncp})). Then one can express the action of the symmetry algebra generators in 
terms of $\hat{x}^\mu$ and derivatives (see \cite{AW,BLS}). 

The action of momentum generators $P_{\mu}$ on classical and quantum fields
$\varphi (x)$ is supposed to be the same 
$$
P_{\mu}\varphi(x)=i\frac{\partial}{\partial x^{\mu}}\varphi(x).
$$
However, in classical theory fields are given by different smooth functions
as elements of ${\mathcal C}(\mathcal M)$ with Fourier expansion (\ref{ft}) 
and the generators $P_{\mu}$ are realized as partial derivatives
$P_{\mu}=i\partial/\partial x^{\mu}$. In quantum theory $\varphi (x)$ and
$P_{\mu}$ are fixed operators as elements of the algebra of observables 
$\mathcal A$. The action of $P_{\mu}$ on $\varphi(x)$ is defined by the
commutator  
$$
P_{\mu}\cdot \varphi (x)=[P_{\mu},\varphi(x)],
$$
and having in mind the expansion of $\varphi(x)$ in terms of the creation
and annihilation  operators $a(k)$, $a^{\dagger}(p)$, one gets 
\begin{equation}\label{Pa}
[P_{\mu},a(k)]=-k_{\mu}a(k).
\end{equation}
We could apply the twisting of the Poincar\'e algebra 
with the action (\ref{Pa}) after quantizing field theory \cite{CPT}. 

Using the twist element 
$\mathcal F=\exp (\frac i2\theta^{\mu \nu} P_{\mu}\otimes P_{\nu})$, 
we have to change product of observables according to the general rule 
\begin{gather}
a*b=m\circ (e^{-\frac i2\theta^{\mu\nu}P_{\mu}\otimes P_{\nu}})(a\otimes b)
\nonumber\\
=m\circ \bigg(\sum\limits^{\infty}_{n=0} \frac{1}{n!}
\bigg(\frac{-i}{2}\bigg)^n\prod\limits^{n}_{j=1}\theta^{\mu_j\nu_j} ad
P_{\mu_j}\otimes ad P_{\nu_j}\bigg) (a\otimes b).
\end{gather}
Hence, the twisted products of the creation and annihilation 
operators are related with the standard products as follows  
\begin{equation}\label{aapr}
a(k)*a(p)=a(k)a(p)e^{-i\theta(k,p)}
\end{equation}  
\begin{equation}\label{aa+pr}
a(k)*a^{\dagger}(p)=a(k)a^{\dagger}(p)e^{i\theta(k,p)}.
\end{equation}
One can see that the $\ast$-products of the exponentials (\ref{newpr}) and 
annihilation or creation operators (\ref{aapr}) have the same factors. 
Being expressed in terms of the twisted product, the commutation
relations are 
\begin{gather}
a(k)*a(p)=a(p)*a(k)e^{-2i\theta(k,p)}  \nonumber\\ 
\label{zf}
a(k)*a^{\dagger}(p)-e^{2i\theta(k,p)}a^{\dagger}(p)*a(k)=\delta (k-p),
\end{gather}
where $\theta(k,p)=-\theta(p,k)=\frac 12\theta^{\mu\nu}k_{\mu}p_{\nu}$. 
The relations (\ref{zf}) reproduce a scalar Zamolodchikov-Faddeev algebra. 
Noncommutativity of both coordinates $x^\mu$ and the operators 
$a(k)$, $a^{\dagger}(k)$ was introduced in \cite{BPQ} with a phase 
in (\ref{zf}) inconsistent with the twisted Poincare algerba 
(see detailed explanation in \cite{AT}). 

 The parameters $\Lambda^\mu_{\,\,\nu}(\omega),$ $a^\mu$
of the global Poincar\'e transformations generate the algebra of
functions $F(G)$ on the Poincar\'e group  $G$. This commutative algebra
$F(G) \simeq ({\mathcal U}({\mathcal P}))^*$ is dual to
${\mathcal U}({\mathcal P})$, and after twisting
${\mathcal U}({\mathcal P})$ the product of the dual Hopf algebra
$({\mathcal U}({\mathcal P}))^*$ is changed.

An important object connecting a pair of dual Hopf algebras is the
canonical element (a bicharacter) \cite{Dr1}
$$
{\mathcal T} = \sum e_k \otimes e^k,
\quad  e_k \in {\mathcal H}^*, \; e^k \in {\mathcal H}, \; 
\langle e_k, e^m  \rangle= \delta_k^m ,
$$
where $e_k$ and $e^m$ are dual linear bases of ${\mathcal H}^*$ and
${\mathcal H}$. Here we have (using a short hand notation 
$a^{\mu} \otimes P_{\mu} := a^{\mu}P_{\mu}$ etc)
$$
 {\mathcal T} = \exp(i a^{\mu}P_{\mu})\exp(i\omega^{\mu \nu}M_{\mu \nu}).
$$
In the case of the twist (\ref{abt}) the
generators $\omega^{\mu \nu}$ or $\Lambda^\mu_{\,\,\nu}(\omega)$
are the same (commutative), but the $a^{\mu}$ become noncommutative 
 (see \cite{O,GG,K}),
\begin{equation}\label{nca}
[a^{\mu}, a^{\nu}]= i\theta^{\mu \nu}-
i\Lambda^\mu_{\,\,\alpha}\Lambda^\nu_{\,\,\beta}\theta^{\alpha \beta}.
\end{equation}
This can be obtained from the RTT-relations \cite{FRT} using the matrix
representation of ${\mathcal U}({\mathcal P})$ and the $R$-matrix,
or from the general recipe (\ref{tp}) using
the ${\mathcal U}({\mathcal P})$-bimodule structure of
$({\mathcal U}({\mathcal P}))^*$.
Due to the commutativity of $\Lambda(\omega)$, in the representations 
$V$ of $({\mathcal U}_t({\mathcal P}))^*$
with $\Lambda^\mu_{\,\,\alpha}=\delta^\mu_{\,\,\alpha},$ 
the generators $a^{\mu}$ are commutative in such $V.$ However, if 
$\Lambda^\mu_{\,\,\alpha} \neq \delta^\mu_{\,\,\alpha},$ then 
$a^{\mu}$ are noncommutative and can not be put to zero. 

The transformation of the coordinates $x^\mu$ is given by the coaction
$\delta : C_\theta\rightarrow F_\theta(G) \otimes C_\theta$,
\begin{equation}\label{trcoor}
\tilde{x}^\mu : = \delta(x^\mu)=
\Lambda^\mu_{\,\,\alpha}\otimes x^\alpha+a^\mu \otimes 1\,.
\end{equation}
The transformed generators satisfy the same relations,
$[\tilde{x}^{\mu}, \tilde{x}^{\nu}]= i\theta^{\mu \nu}.$ Hence one can
conclude that the noncommutative space-time (\ref{cr01}) is invariant under
the twisted Poincar\'e algebra  ${\mathcal U}_t({\mathcal P}).$ The action 
of the algebra generators can be integrated 
$$
\frac{df}{d\tau} = \alpha_j Y_j f \; \rightarrow \; 
f(\tau) = \exp(\alpha_j Y_j \tau) f(0). 
$$
However, to have a similar action on the star product of two elements 
$f,\, g$ the parameters 
$\alpha_j$ have to be noncommutative (see (\ref{nca})). 

Tensoring\, two copies of\, the NC\, space-time algebra,
\,\,${\mathcal C}_{\theta}\,\otimes\, {\mathcal C}_{\theta}$\,\,
with\, generators
$x^{\mu}_1=x^{\mu}\otimes 1$ and
$x^{\nu}_2=1\otimes x^{\nu}$, one gets their commutation relations
according to (\ref{ttp}) with R-matrix
${\mathcal R}=\exp\left(-i\theta^{\mu \nu}P_{\mu}\otimes P_{\nu}\right)$
\cite{K}:
\begin{equation}\label{cr2}
\begin{split}
x^{\mu}_1x^{\nu}_2-x^{\nu}_2x^{\mu}_1:=x^{\mu}\otimes x^{\nu}
-(1\otimes x^{\nu})(x^{\mu}\otimes 1)\qquad\qquad\qquad\qquad\qquad\qquad
\\
=x^{\mu}\otimes x^{\nu}-
\sum_{k=0}^{\infty}\frac{(i)^k}{k!}\prod_{j=1}^{k}\theta^{\mu_j, \nu_j}
\left(\partial_{\nu_1}\ldots\partial_{\nu_k}x^{\mu}\right) \otimes
\left(\partial_{\mu_1}\ldots\partial_{\mu_k}x^{\nu}\right)
\qquad\\
=x^{\mu}\otimes x^{\nu}-x^{\mu}\otimes x^{\nu}-i\theta^{\nu \mu}=
i\theta^{\mu \nu}.
\end{split}
\end{equation}
This property results in an extra factor in the Fourier transform
of the vacuum expectation value
$\langle \varphi(x_1) \ast \varphi(x_2) \ast \cdots
\ast \varphi(x_n)\rangle$ of quantum fields \cite{O}.
Application of the abelian twist to gauge field theories are also 
considered (see \cite{CT} and references therein).

Similar arguments as in \cite{CK} can be applied in the case of (extended) 
supersymmetry and of the Poincar\'e superalgebra $s{\mathcal P}$ 
with additional supercharges (odd generators)
$Q_{\alpha}, \bar{Q}_{\dot{\beta}}$ to get a noncommutative superspace as
in \cite{S}. The Poincar\'e Lie superalgebra commutation relations
(the commutators below are $Z_2$-graded, i.e.\ if both elements are odd
it is the anticommutator) are
\begin{align*}
[P_\mu, Q_\alpha] & = 0,&
[M_{\mu \nu}, Q_{\alpha}] & = i (\sigma_{\mu\nu})_\alpha^{\,\,\beta} Q_{\beta},
\\[4pt]
[Q_{\alpha}, Q_{\beta}] & = 0,&
[M_{\mu \nu}, {\bar{Q}}_{\dot{\beta}}] & =
i ({\bar{\sigma}}_{\mu\nu})_{\dot{\beta}}^{\,\,\dot{\alpha}}
\bar{Q}_{\dot{\alpha}},
\\[4pt]
[\bar{Q}_{\dot{\alpha}},\bar{Q}_{\dot{\beta}}] & = 0,&
[Q_{\alpha}, \bar{Q}_{\dot{\beta}}] & =
2 \sigma_{\alpha \dot{\beta}}^\mu P_\mu.
\end{align*}
The generators $P_\mu, Q_{\alpha}$ define an abelian (supercommutative)
subalgebra, and abe\-li\-an twists depending  on odd
generators 
can be constructed as in the non-graded case, e.g. 
%
\begin{equation}\label{st}
{\mathcal F} = exp ( C^{\alpha \beta} Q_{\alpha} \otimes Q_{\beta}) =  
1 + C^{\alpha \beta} Q_{\alpha} \otimes Q_{\beta} - 
\det C \, Q_1 Q_2 \otimes Q_1 Q_2 , 
\end{equation}
%
with symmetric matrix $C^{\alpha \beta} = C^{\beta \alpha}$.
The exponent reproduces a Poisson tensor defining superbrackets
(see e.g.\ \cite{ILZ}), and can be used to construct
noncommutative superspace preserving super-Poincar\'e covariance 
\cite{KoSa,Z}.

The algebraic sector of the twisted Hopf superalgebra
${\mathcal U}_t(s {\mathcal P})$ is not changed, as well as the coproduct
of the abelian subalgebra of (super)translations
with the generators $P_\mu, Q_{\alpha}$. However, the
coproducts of $M_{\mu \nu}$ and $\bar{Q}_{\dot{\beta}}$ are changed:
\[
\begin{split}
\Delta_t (M_{\mu \nu}) & = {\mathcal F} \Delta (M_{\mu \nu})
{\mathcal F}^{-1} \\[4pt]
& = \Delta(M_{\mu \nu}) - i\{C^{\alpha \beta}(\sigma_{\mu
  \nu})^\gamma_\alpha + C^{\alpha \gamma}(\sigma_{\mu \nu})^\beta_\alpha\}
Q_\gamma \otimes Q_\beta,\\[-9pt]~
\end{split}
\]
\[
\Delta_t(\bar{Q}_{\dot{\gamma}})=\Delta(\bar{Q}_{\dot{\gamma}}) +
2C^{\alpha \beta}\sigma^{\mu}_{\alpha \dot{\gamma}}
\left( Q_{\beta}\otimes P_{\mu} - P_{\mu}\otimes Q_{\beta}\right).\qquad\quad
\]
The standard realization of the supercharges
\[
Q_\alpha= \partial/\partial \theta^\alpha - i 
 \sigma^\mu_{\alpha \dot{\beta}} \bar{\theta}^{\dot{\beta}} \partial/\partial x^\mu ,
 \qquad 
\bar{Q}_{\dot{\beta}} = \partial/\partial \bar{\theta}^{\dot{\beta}} 
 - i \theta^\alpha \sigma^\mu_{\alpha \dot{\beta}} \partial/\partial x^\mu 
\]
yields noncommutative generators $x^\mu, \theta^\alpha, \bar{\theta}^{\dot{\beta}}$ of
Minkowski superspace $s{\mathcal M}_t$,
\[
[\theta^\alpha, \theta^\beta] = 2 C^{\alpha\beta},\qquad
[x^\mu,\theta^\alpha] = 2i C^{\alpha\beta} \sigma^\mu_{\beta \dot{\gamma}}
\bar{\theta}^{\dot{\gamma}},\qquad
[x^\mu, x^\nu] = 2 C^{\alpha \beta} \sigma^\mu_{\alpha \dot{\gamma}}
\sigma^\nu_{\beta \dot{\delta}} \bar{\theta}^{\dot{\gamma}}
\bar{\theta}^{\dot{\delta}}.
\]
These commutators are consequences of the star products 
$a \ast b = m \circ (exp (- C^{\alpha \beta} Q_{\alpha} 
\otimes Q_{\beta})) (a \otimes b)$
\[
\theta^\alpha \ast \theta^\beta =\theta^\alpha \cdot \theta^\beta + C^{\alpha\beta} , 
\qquad x^\mu \ast \theta^\alpha = x^\mu \cdot \theta^\alpha + 
iC^{\alpha\beta} \sigma^\mu_{\beta \dot{\gamma}} \bar{\theta}^{\dot{\gamma}} , 
\]
\[
x^\mu \ast x^\nu = x^\mu \cdot x^\nu + 
C^{\alpha\beta} \sigma^\mu_{\alpha \dot{\gamma}} \bar{\theta}^{\dot{\gamma}}
\sigma^\nu_{\beta \dot{\delta}} \bar{\theta}^{\dot{\delta}}. 
\]
As in the previous case one can consider realization of the $\ast$-multiplication 
of the standard/usual elements of the algebra of functions $\varphi (x, \theta,  \bar{\theta})$ 
by $ \theta^\alpha \ast $ or by $x^\mu \ast $ as an action by the operators 
\[
\theta^\alpha \ast  = \theta^\alpha \cdot   + 
C^{\alpha \beta} (\partial/\partial \theta^\beta - 
i\sigma^\nu_{\beta \dot{\beta}} \bar{\theta}^{\dot{\beta}} \partial/\partial x^\nu),  
\qquad x^\mu \ast   = x^\mu \cdot   + 
i C^{\alpha\beta} \sigma^\mu_{\alpha \dot{\gamma}} \bar{\theta}^{\dot{\gamma}} 
( \partial/\partial \theta^\beta -i\sigma^\nu_{\beta \dot{\delta}} \bar{\theta}^{\dot{\delta}} 
\partial/\partial x^\nu). 
 \]

It is important to point out that generators (parameters) of the deformed
Poincar\'e supergroup $({\mathcal U}_t(s {\mathcal P}))^*$ : 
$\omega^{\mu \nu}, b^\mu, \lambda^\alpha, \bar{\lambda}^{\dot{\beta}}$
dual to $M_{\mu \nu}, P_\mu, Q_\alpha, \bar{Q}_{\dot{\beta}}$ 
will be not supercommutative. However, their commutation relations will be
different from those of $s{\mathcal M}_t$.

Representing the canonical element ${\mathcal T}$ of the twisted
Poincar\'e superalgebra ${\mathcal U}_t(s {\mathcal P})$ and its
dual quantum Poincar\'e supergroup $({\mathcal U}_t(s {\mathcal P}))^*$ in the form
%
\begin{equation}\label{ce}
{\mathcal T} = 
\exp (i\bar{\lambda}^{\dot{\alpha}} \bar{Q}_{\dot{\alpha}})
\exp (i\lambda^\alpha Q_\alpha)
\exp (ib^\mu P_\mu) \exp (i\omega^{\mu \nu} M_{\mu \nu}),
\end{equation}
%
one can get e.g.\ from the RTT-relation \cite{FRT} that 
$\omega^{\mu \nu}, \bar{\lambda}^{\dot{\alpha}}$  are 
supercommutative while 
$\lambda^\alpha$ and $b^\mu$ do not super-commute e.g. 
\[
[\lambda^\alpha, \lambda^\beta] = 2C^{\alpha \beta} - 
2(S(\omega))^\alpha_{\,\,\gamma}(S(\omega))^\beta_{\,\,\delta}C^{\gamma \delta}. 
\]
Let us point out that with the form (\ref{ce}) of the canonical element 
the generators $b^\mu$ are not hermitian.
The twist (\ref{st}) also does not preserve the standard $\ast$-operation of 
the Poincar\'e superalgebra. 
One can also use the twist (\ref{st}) and coproducts of the generators 
$\omega^{\mu \nu}, b^\mu, \lambda^\alpha, \bar{\lambda}^{\dot{\beta}}$ 
\[
\Delta (\omega^{\mu \nu}) = {\mathcal D}^{\mu \nu} (\omega \otimes 1, 1 \otimes \omega), 
\]
\[
\Delta (\lambda^\alpha) = \lambda^\alpha \otimes 1 + 
(S(\omega))^\alpha_{\,\,\beta} \otimes \lambda^\beta, 
\Delta (\bar{\lambda}^{\dot{\alpha}}) = \bar{\lambda}^{\dot{\alpha}} \otimes 1 + 
(\bar{S}(\omega))^{\dot{\alpha}}_{\,\,\dot{\delta}} \otimes \bar{\lambda}^{\dot{\delta}},
\]
\[
\Delta (b^\mu) = b^\mu \otimes 1 + \Lambda (\omega)^{\mu}_\nu \otimes b^\nu - 2i
(\lambda^\alpha (\bar{S}(\omega))^{\dot{\alpha}}_{\,\,\dot{\delta}} \otimes \bar{\lambda}^{\dot{\delta}})
\sigma^{\mu}_{\alpha \dot{\alpha}}
\]
to get corresponding star products. 
${\mathcal D}^{\mu \nu} (\omega \otimes 1, 1 \otimes \omega)$ is the BCH series of the Lorentz 
algebra and 
$\Lambda (\omega), \bar{S}(\omega), (S(\omega))^\beta_{\,\,\delta}$
are the Lorentz transformation matrices acting on the vector and Weyl 
spinor indices. The commutation relations of the NC superspace
$s{\mathcal M}_t$ are invariant with respect to the twisted 
Poincar\'e superalgebra ${\mathcal U}_t(s{\mathcal P})$ and supergroup. 

Some consequences of the twisted Poincar\'e supersymmetry applied to 
the Wess-Zumino model are discussed in \cite{ACNT}. 

\medskip 

{\bf Acknowledgment} 

The author wants to thank the organizers of the conference "Non-Commutative Geometry 
and Representation Theory in Mathematical Physics", Karlstad, Sweden, 2004, and 
Masud Chaichian and Anca Tureanu for useful discussions.  

\medskip

 \def\czjp  {Czech.\,J.\,Phys.}
 \def\jhep  {J.\,High\,En\-er\-gy\,Phys.}
 \def\leni  {Le\-nin\-grad\,Math.\,J.}
 \def\nupb  {Nucl.\,Phys.~B}
 \def\phlb  {Phys.\,Lett.~B}
 \def\phrp  {Phys.\,Rep.}
 \def\remp  {Rev.\,Mod.\,Phys.}
 \def\slnp  {Sprin\-ger\,Lec\-ture\,No\-tes\,in\,Phy\-sics}

\medskip

\end{document}